\documentclass[twoside,twocolumn,9pt]{article}
\usepackage{extsizes}
\usepackage[super,sort&compress,comma]{natbib} 
\usepackage[version=3]{mhchem}
\usepackage[left=1.5cm, right=1.5cm, top=1.785cm, bottom=2.0cm]{geometry}
\usepackage{balance}
\usepackage{mathptmx}
\usepackage{sectsty}
\usepackage{graphicx} 
\usepackage{lastpage}
\usepackage[format=plain,justification=justified,singlelinecheck=false,font={stretch=1.125,small,sf},labelfont=bf,labelsep=space]{caption}
\usepackage{float}
\usepackage{fancyhdr}
\usepackage{fnpos}
\usepackage[english]{babel}
\addto{\captionsenglish}{%
  
}
\usepackage{array}
\usepackage{droidsans}
\usepackage{charter}
\usepackage[T1]{fontenc}
\usepackage[usenames,dvipsnames]{xcolor}
\usepackage{setspace}
\usepackage[compact]{titlesec}
\usepackage{hyperref}

\usepackage{epstopdf}

\definecolor{cream}{RGB}{222,217,201}

\begin{document}

\pagestyle{fancy}
\thispagestyle{plain}
\fancypagestyle{plain}{
\renewcommand{\headrulewidth}{0pt}
}

\makeFNbottom
\makeatletter
\renewcommand\LARGE{\@setfontsize\LARGE{15pt}{17}}
\renewcommand\Large{\@setfontsize\Large{12pt}{14}}
\renewcommand\large{\@setfontsize\large{10pt}{12}}
\renewcommand\footnotesize{\@setfontsize\footnotesize{7pt}{10}}
\makeatother

\renewcommand{\thefootnote}{\fnsymbol{footnote}}
\renewcommand\footnoterule{\vspace*{1pt}%
\color{cream}\hrule width 3.5in height 0.4pt \color{black}\vspace*{5pt}} 
\setcounter{secnumdepth}{5}

\makeFNbottom
\makeatletter
\renewcommand\LARGE{\@setfontsize\LARGE{15pt}{17}}
\renewcommand\Large{\@setfontsize\Large{12pt}{14}}
\renewcommand\large{\@setfontsize\large{10pt}{12}}
\renewcommand\footnotesize{\@setfontsize\footnotesize{7pt}{10}}
\renewcommand\scriptsize{\@setfontsize\scriptsize{7pt}{7}}
\makeatother

\renewcommand{\thefootnote}{\fnsymbol{footnote}}
\renewcommand\footnoterule{\vspace*{1pt}%
\color{cream}\hrule width 3.5in height 0.4pt \color{black} \vspace*{5pt}} 
\setcounter{secnumdepth}{5}

\makeatletter 
\renewcommand\@biblabel[1]{#1}            
\renewcommand\@makefntext[1]%
{\noindent\makebox[0pt][r]{\@thefnmark\,}#1}
\makeatother 
\renewcommand{\figurename}{\small{Fig.}~}
\sectionfont{\sffamily\Large}
\subsectionfont{\normalsize}
\subsubsectionfont{\bf}
\setstretch{1.125} 
\setlength{\skip\footins}{0.8cm}
\setlength{\footnotesep}{0.25cm}
\setlength{\jot}{10pt}
\titlespacing*{\section}{0pt}{4pt}{4pt}
\titlespacing*{\subsection}{0pt}{15pt}{1pt}

\fancyfoot{}
\fancyfoot[LO,RE]{\vspace{-7.1pt}\includegraphics[height=9pt]{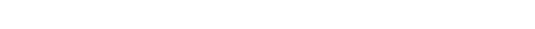}}
\fancyfoot[CO]{\vspace{-7.1pt}\hspace{13.2cm}\includegraphics{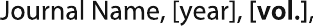}}
\fancyfoot[CE]{\vspace{-7.2pt}\hspace{-14.2cm}\includegraphics{head_foot/RF}}
\fancyfoot[RO]{\footnotesize{\sffamily{1--\pageref{LastPage} ~\textbar  \hspace{2pt}\thepage}}}
\fancyfoot[LE]{\footnotesize{\sffamily{\thepage~\textbar\hspace{3.45cm} 1--\pageref{LastPage}}}}
\fancyhead{}
\renewcommand{\headrulewidth}{0pt} 
\renewcommand{\footrulewidth}{0pt}
\setlength{\arrayrulewidth}{1pt}
\setlength{\columnsep}{6.5mm}
\setlength\bibsep{1pt}

\makeatletter 
\newlength{\figrulesep} 
\setlength{\figrulesep}{0.5\textfloatsep} 

\newcommand{\topfigrule}{\vspace*{-1pt}%
\noindent{\color{cream}\rule[-\figrulesep]{\columnwidth}{1.5pt}} }

\newcommand{\botfigrule}{\vspace*{-2pt}%
\noindent{\color{cream}\rule[\figrulesep]{\columnwidth}{1.5pt}} }

\newcommand{\dblfigrule}{\vspace*{-1pt}%
\noindent{\color{cream}\rule[-\figrulesep]{\textwidth}{1.5pt}} }

\makeatother


\twocolumn[
  \begin{@twocolumnfalse}
{\includegraphics[height=30pt]{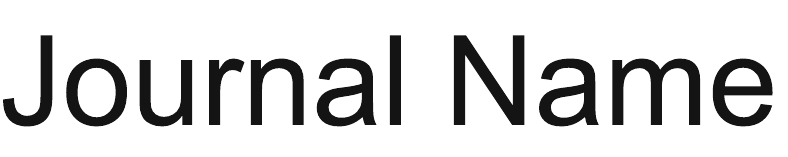}\hfill\raisebox{0pt}[0pt][0pt]{\includegraphics[height=55pt]{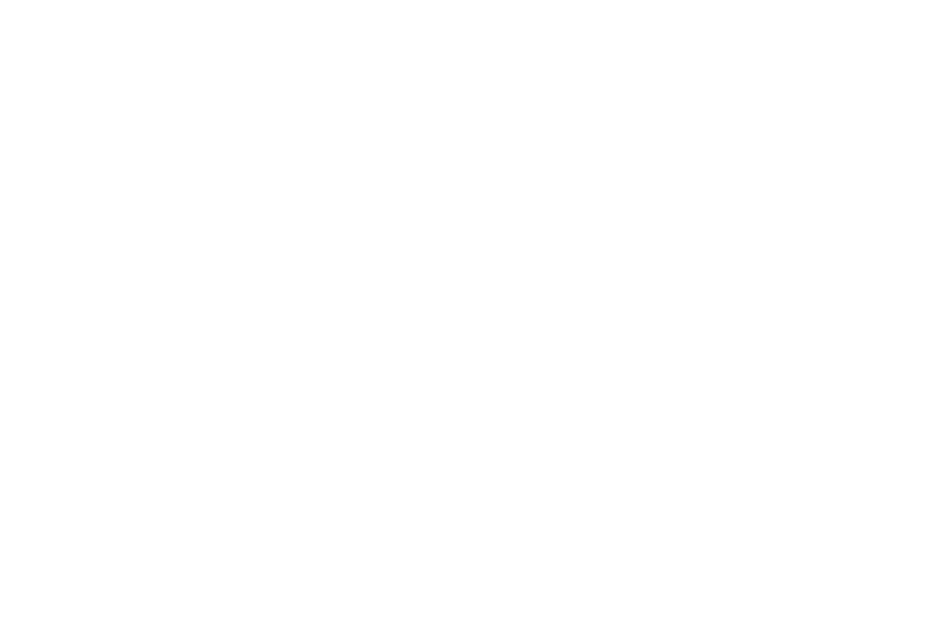}}\\[1ex]
\includegraphics[width=18.5cm]{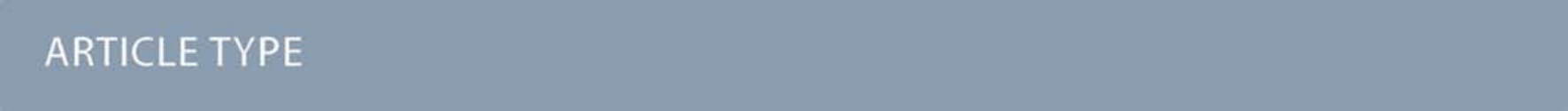}}\par
\vspace{1em}
\sffamily
\begin{tabular}{m{4.5cm} p{13.5cm} }

\includegraphics{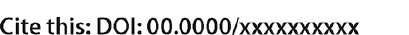} & \noindent\LARGE{\textbf{Molecular theory of electrostatic collapse of dipolar polymer gels$^\dag$}} \\
\vspace{0.3cm} & \vspace{0.3cm} \\

 & \noindent\large{Yury A. Budkov$^{\ast}$\textit{$^{a,}$}\textit{$^{b}$}, Nikolai N. Kalikin\textit{$^{b}$} and Andrei L. Kolesnikov\textit{$^{c}$}} \\

\includegraphics{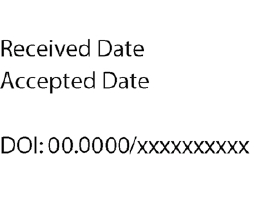} & \\

\end{tabular}

 \end{@twocolumnfalse} \vspace{0.6cm}
]

\renewcommand*\rmdefault{bch}\normalfont\upshape
\rmfamily
\section*{}
\vspace{-1cm}


\footnotetext{\textit{$^{a}$School of Applied Mathematics, HSE University, Tallinskaya st. 34, 123458 Moscow, Russia. }}
\footnotetext{\textit{$^{b}$G.A. Krestov Institute of Solution Chemistry of the Russian Academy of Sciences, Academicheskaya st., 1, 153045 Ivanovo, Russia.}}
\footnotetext{\textit{$^{c}$Institut f\"{u}r Nichtklassische Chemie e.V., Permoserstr. 15, 04318 Leipzig, Germany}}
\footnotetext{$^{\ast}$ ybudkov@hse.ru}

\footnotetext{\dag~Electronic Supplementary Information (ESI) available: [details of any supplementary information available should be included here]. See DOI: 00.0000/00000000.}




\sffamily{\bf{We develop a new quantitative molecular theory of liquid-phase dipolar polymer gels. We model monomer units of the polymer network as a couple of charged sites separated by a fluctuating distance. For the first time, within the random phase approximation, we have obtained an analytical expression for the electrostatic free energy of the dipolar gel. Depending on the coupling parameter of dipole-dipole interactions and the ratio of the dipole length to the subchain Kuhn length, we describe the gel collapse induced by electrostatic interactions in the good solvent regime as a first-order phase transition. This transition can be realized at reasonable physical parameters of the system (temperature, solvent dielectric constant, and dipole moment of monomer units). The obtained results could be potentially used in modern applications of stimuli-responsive polymer gels and microgels, such as drug delivery, nanoreactors, molecular uptake, coatings, superabsorbents, {\sl etc.}}}\\


\rmfamily 

\pagestyle{fancy}
In recent years a considerable number of papers have been devoted to the description of the stimuli-responsive polymer systems and their potential application in the field of externally controlled drug delivery, sensing and biosensing, artificial muscles and actuators, {\sl etc} \cite{bawa2009stimuli,wei2017stimuli,cabane2012stimuli}. One of the phenomena underlying such smart behavior is the coil-globule (CG) conformation transition of a single polymer chain in the solvent media under different external stimuli \cite{khokhlov1994statistical,lifshitz1978some,budkov2017flory,budkov2018models}. 
Stimuli-responsive polymer gels and microgels are now indispensable to a wide range of industrial applications, such as drug delivery, coatings, adsorbents, molecular chemical reactors, {\sl etc} (see, for instance, \cite{stuart2010emerging,goponenko2016role,de2005stimuli,angioletti2015theory,kanduvc2020modeling}). The existence of the volume phase transition of the polymer network was first predicted theoretically \cite{duvsek1968transition} as an analogy with the CG transition, occurring in the dilute polymer solutions, and later was observed experimentally \cite{tanaka1978collapse}. In the following years an extensive investigation, both theoretical and experimental, was carried out to examine the volume phase transition of polymer networks, including polyelectrolyte ones \cite{hirotsu1987volume,hirokawa1984volume,tanaka1980phase,khokhlov1980swelling,khokhlov1993conformational,khokhlov1994polyelectrolyte,sing2013effect,philippova2013new,rumyantsev2016polyelectrolyte,polotsky2013collapse,molotilin2018star}. Special attention should be paid to the work \cite{zaroslov1999change}, where the authors experimentally studied the collapse of a polyelectrolyte polymer gel in a poor solvent regime. The authors discussed an influence of additional dipole-dipole attraction of the ionic pairs formed on the polymer backbone due to the counterion condensation on the collapse of gel.   

\begin{figure}
\center{\includegraphics[height=3.5 cm]{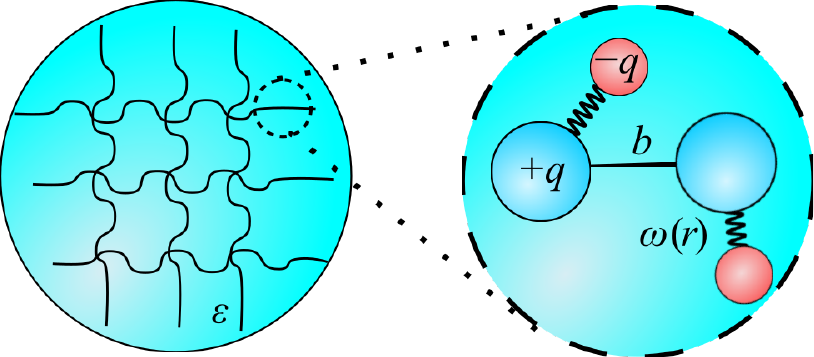}}
\caption{Schematic representation of the dipolar gel model.}
\label{fig1}
\end{figure}

There is a number of polymer systems, the thermodynamic behavior of which is strongly dependent on the dipole-dipole interactions of highly polar monomer units\cite{martin2016statistical}. Among them are ionomers \cite{zhang2014perspective}, dielectric elastomers \cite{suo2010theory}, zwitterionic polymers \cite{lowe2002synthesis,kudaibergenov2006polymeric,lei2018zwitterionic}, polymeric ionic liquids \cite{kohno2015thermoresponsive}, to name a few. One of the brightest effects driven by the dipole-dipole interactions is the CG transition of a single dipolar polymer chain \cite{budkov2017polymer}. The latter can occur in polyelectrolyte solutions with low-polar solvents \cite{schiessel1998counterion,cherstvy2010collapse,gordievskaya2018interplay}, where the monomer units and counterions form ionic pairs, as well as in dilute solutions of polyzwitterionic macromolecules (betaines), where the monomer units carry two oppositely charged ionic groups \cite{kumar2009theory}. Recently, this fascinating phenomenon has been extensively studied within the molecular dynamics simulations \cite{gordievskaya2018interplay,gordievskaya2019conformational}. 

In this short communication, we report a new quantitative molecular theory of phase behavior of the polymer liquid-phase gels with highly polar monomer units. We describe the polymer network behavior within a wide range of microscopic parameters underlying the gel collapse, induced by electrostatic interactions, in the good solvent regime.

Let us consider a polymer gel -- polymer network, immersed in a polar solvent with a certain dielectric constant $\varepsilon$, whose monomer units carry two oppositely charged sites with $\pm q$ charges separated by a fluctuating distance, described by the probability distribution function, $\omega({\bf r})$ (see Fig. 1). The total free energy of such a {\sl dipolar} gel in the equilibrium state can be written as a sum of three basic terms
\begin{equation}
F_{tot}=F_{el}+F_{vol}+F_{cor},
\end{equation}
where $F_{el}$ is the elastic free energy of the network, $F_{vol}$ is the contribution of the short-range volume (excluded volume and dispersion) interactions of the monomer units, and $F_{cor}$ is the contribution of their electrostatic correlations, attributed to the charged sites on each monomer unit. According to the conventional concept \cite{shibayama1993volume,khokhlov1993conformational} adopted in the classical theory of gels, we can estimate elastic free energy as the free energy of $n$ independent Gaussian subchains, i.e. by the following interpolation formula  \cite{grosberg1992quantitative,brilliantov1998chain,budkov2016new,budkov2018models} 
\begin{equation}
F_{el}=\frac{9}{4}n k_{B}T\left((V/V_0)^{2/3}+(V/V_0)^{-2/3}\right),
\end{equation}
where $V$ is the gel volume and $V_0$ is the gel volume at the reference state corresponding to the $\theta$-temperature \cite{khokhlov1980swelling,tanaka1978collapse} at which the subchains behave as the ideal Gaussian coils; $k_{B}$ is the Boltzmann constant and $T$ is the temperature. The contribution of the volume interactions of the monomer units can be described by the well-known Flory-Huggins (FH) mean-field approximation \cite{shibayama1993volume,khokhlov1994statistical,de1979scaling}
\begin{equation}
F_{vol}=\frac{Vk_{B}T}{v}\left(\left(1-\Phi\right)\ln\left(1-\Phi\right)+\Phi-\chi \Phi^2\right),
\end{equation}
where $\Phi=n m v/V=cv$ ($c=n m/V$ is the concentration of the monomer units) is the volume fraction of the polymer, $v$ is the effective volume, occupied by the monomer unit, $m$ is the average polymerization degree of the subchain. The electrostatic (correlation) free energy can be calculated within the random phase approximation (RPA) \cite{borue1988statistical,budkov2020statistical,budkov2019statistical,budkov2018nonlocal,budkov2019nonlocal} as follows
\begin{equation}
F_{cor}=\frac{Vk_{B}T}{2}\int\frac{d\bf{k}}{(2\pi)^3}\left(\ln\left(1+\frac{\kappa^2(\bf{k})}{k^2}\right)-\frac{\kappa^2(\bf{k})}{k^2}\right),
\end{equation}
where the screening function takes the form (see the Supporting information) 
\begin{equation}
\label{screening_function}
\kappa^2({\bf k})=\kappa_{D}^2(1-\omega({\bf k}))\left(1+\frac{1}{2}\left(S({\bf k})-1\right)(1-\omega({\bf k}))\right),
\end{equation}
where $\kappa_{D}=\left(8\pi q^2c/\varepsilon k_{B}T\right)^{1/2}$ is the inverse Debye length, attributed to the charged groups on the monomer segments and $\omega({\bf k})=\int d{\bf r} e^{-i{\bf k}{\bf r}}\omega({\bf r})$ is the Fourier-image of the probability distribution function (characteristic function). The structure factor of the subchain at $m\gg 1$ can be approximated by the well-known Debye structure factor $S({\bf k})\approx 1+12/(k^2b^2)$ of the Gaussian chain, where $b$ is the Kuhn length of the segment. Using the {\sl ansatz} for the Fourier-image of the probability distribution function of distance between charged centers (characteristic function) $\omega({\bf k})=(1+k^2l^2/6)^{-1}$, where $l$ is the effective dipole length, determining the following probability distribution function \cite{budkov2018nonlocal,budkov2020statistical} $\omega({\bf r})=3/(2\pi l^2 |{\bf r}|)\exp\left[-\sqrt{6}|{\bf r}|/l\right]$, we arrive at the analytical expression
\begin{equation}
\label{Fcor}
F_{cor}=F_{cor}^{(d)}+F_{cor}^{(ch)},
\end{equation}
where the first term is the electrostatic free energy of unbound dipolar monomers, determined by the expression \cite{budkov2020statistical} 
\begin{equation}
\label{dipol}
F_{cor}^{(d)}=-\frac{Vk_{B}T}{l^3}\sigma(y),
\end{equation} 
with the strength of the electrostatic interactions $y=4\pi p^2c/(3 k_{B}T\varepsilon)$ ($p=ql$ is the dipole moment) and auxiliary function 
\begin{equation}
\sigma(y)=\frac{\sqrt{6}}{4\pi}\left(2(1+y)^{3/2}-2-3y\right).
\end{equation}
The second term determines the effect of the polymerization of the dipolar monomer units on the electrostatic free energy and can be written in the following form 
\begin{equation}
\label{Fcor,ch}
F_{cor}^{(ch)}=-\frac{Vk_{B}T}{l^3}\delta(y,\gamma),
\end{equation}
with the geometric parameter $\gamma=\left(l/b\right)^2$ and the auxiliary functions
\begin{equation}
\delta(y,\gamma)=\delta_1(y,\gamma)+\delta_2(y,\gamma),
\end{equation}
\begin{equation}
\delta_1(y,\gamma)=\frac{3\sqrt{6}}{4\pi}\left(1+\frac{y}{2}-\sqrt{1+y}\right)\gamma,
\end{equation}
\begin{multline}
\delta_2(y,\gamma)=\frac{\sqrt{6}\gamma^2y^2}{4\pi(y-\gamma(1+y))^2}\bigg[1-(2y+5)\sqrt{1+y}-\\\frac{3\gamma (1+y)(1-\sqrt{1+y})}{y}+\frac{1}{\sqrt{2}}\bigg(\bigg(2+y+\sqrt{y^2+4y-4\gamma (1+y)}\bigg)^{3/2}+\\\bigg(2+y-\sqrt{y^2+4y-4\gamma (1+y)}\bigg)^{3/2}\bigg)\bigg].
\end{multline}
The analytical expression for the electrostatic free energy, taking into account electrostatic correlations at the many-body level, as the main result of this work, can be used not only for the dipolar polymer gels description, but also for the description of polymer solutions.

It is instructive to discuss how the electrostatic interactions of the monomer units contribute to the second virial coefficient which, as is well known \cite{de1979scaling}, determines the thermodynamic behavior of the polymer gels and polymer solutions at a rather small polymer volume fraction. Using the analytical expression for the electrostatic free energy (see, Supporting information), we obtain the following excess free energy of the network at $\Phi\ll 1$
\begin{equation}
F_{ex}=F_{vol}+F_{cor}\simeq\frac{Vk_{B}T}{2}Bc^2,
\end{equation}
where we have introduced the second virial coefficient
\begin{equation}
\label{second_vir}
B=B_0-\frac{2\sqrt{6}\pi p^4}{3(\varepsilon k_{B}T)^2l^3}g(\gamma).
\end{equation}
The first term $B_0 = (1-2\chi)v$ on the right hand side defines the second virial coefficient of the interaction between the monomer units without dipole moments within the FH theory. The second term describes the contribution of the dipole-dipole interactions of the monomer units to the second virial coefficient; $g(\gamma)=5\gamma/2+8\sqrt{2}\left(\sqrt{1+\sqrt{1+\gamma}}-\sqrt{\gamma(\sqrt{1+\gamma}-1)}\right)/3-13/3$ is the auxiliary monotonically increasing function. Eq. (\ref{second_vir}) shows that polar particles, tied in a linear chain attract each other more strongly than the corresponding freely moving polar monomers. For the disconnected monomers, for which $\gamma=0$ and $g(0)=1$, we arrive at the expression obtained for the first time in Ref.\cite{budkov2018nonlocal}. As is seen, at rather strong electrostatic interactions (low temperatures, a small dielectric constant, and a large dipole moment), the second virial coefficient can be negative even in the good solvent regime ($\chi < 1/2$). In this case, the dipole-dipole attractive interaction prevails over the short-range volume interactions causing the collapse of the dipolar gel (see below).

For the numerical calculations, we introduce the expansion factor $\alpha=(V/V_{0})^{1/3}$ of the polymer network and the electrostatic coupling parameter, $\lambda=4\pi p^2/(3k_{B}T\varepsilon v)$. Taking into account that $\Phi=\Phi_0/\alpha^3$, where $\Phi_{0}$ is the polymer volume fraction in the reference state, we can minimize the total free energy of the gel with respect to $\alpha$ at different $\lambda$, $\gamma$, and $\Phi_0$. In what follows, we assume that $v/l^3=1$. We consider only the good solvent regime, assuming the FH parameter $\chi=0.25$. Only in this regime we can expect a nontrivial phase behavior of the dipolar polymer gel. As we have already pointed out above, at sufficiently strong electrostatic interactions of the monomer units, the second virial coefficient becomes negative, resulting in polymer network shrinkage. However, to describe the thermodynamics of the shrunk gel, it is necessary to take into account electrostatic correlations on the level higher than the pairwise ones, i.e. use the analytical expressions (\ref{Fcor}), (\ref{dipol}), (\ref{Fcor,ch}) for the electrostatic free energy. It is interesting to study this effect more carefully.

\begin{figure}
\center{\includegraphics[height=7.1 cm]{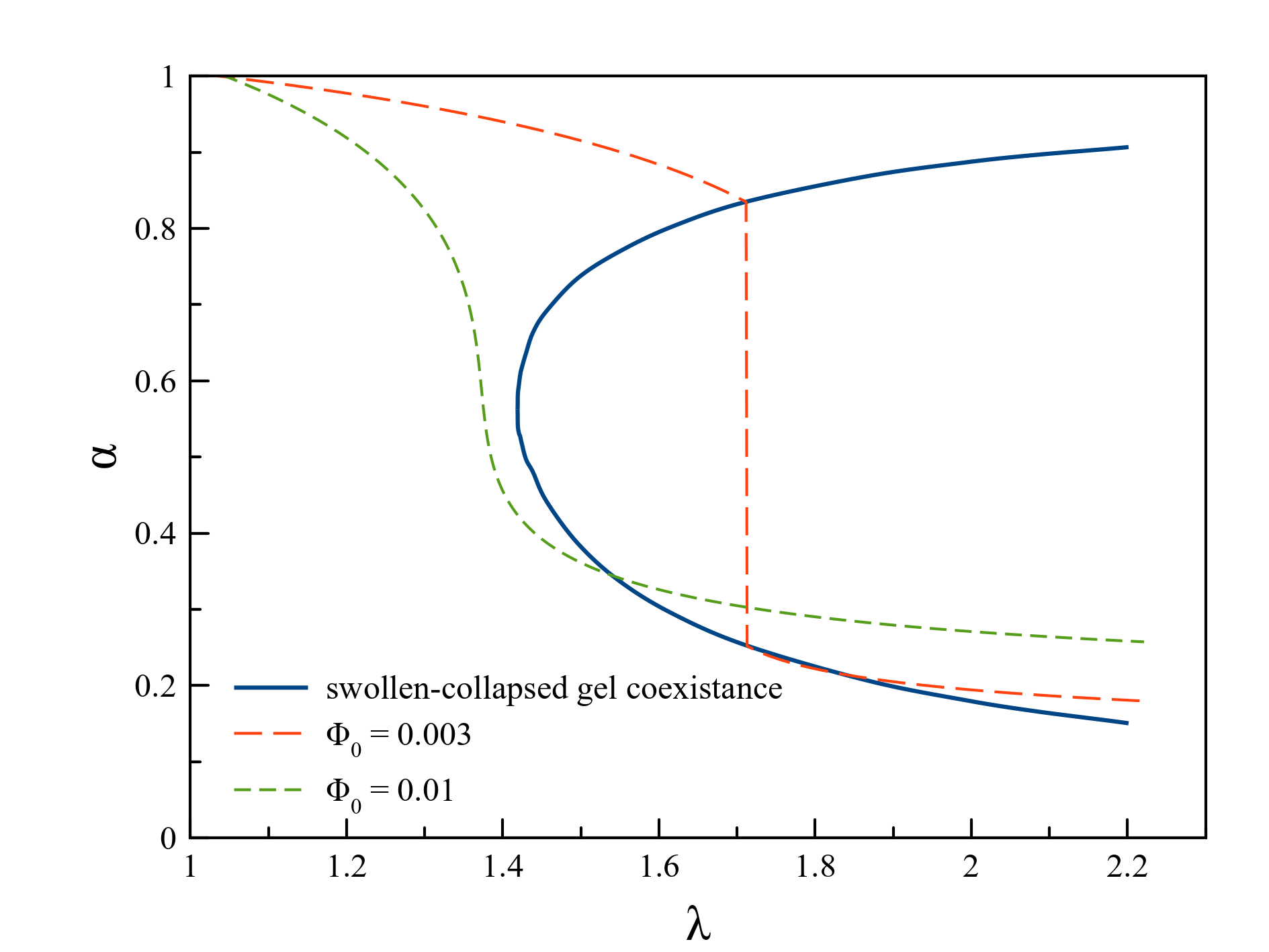}}
\caption{The dependencies of the expansion factor, $\alpha$, on the coupling parameter, $\lambda$, plotted for different reference volume fractions, $\Phi_0$ ($\approx 10^{-3} - 8\times 10^{-3}$). The data are shown for $\chi=0.25$, $v/l^3=1$, $\gamma=1$, $m=500$.}
\label{Fig2}
\end{figure}

Fig. \ref{Fig2} shows the dependencies of the expansion factor $\alpha$ on the coupling parameter $\lambda$, plotted for different reference volume fractions $\Phi_0$ and fixed geometric parameter $\gamma=1$, obtained from the total free energy minimization. As is seen, at sufficiently small $\Phi_0$ values, an increase in the coupling parameter leads to a jump-like collapse of the polymer network, i.e. as a first-order phase transition. However, when the volume fraction exceeds a certain critical value, the transition proceeds smoothly. We would like to stress, however, it is unlikely that the predicted jump-like transition can be observed in real polymer gels due to extremely small corresponding reference volume fractions $\Phi_0$. Thus, in reality, one can expect to achieve the transition in the form of the smooth gel collapse at sufficiently large volume fractions, $\Phi_0$, according to Fig. \ref{Fig2}. 

\begin{figure}
\center{\includegraphics[height=7.1 cm]{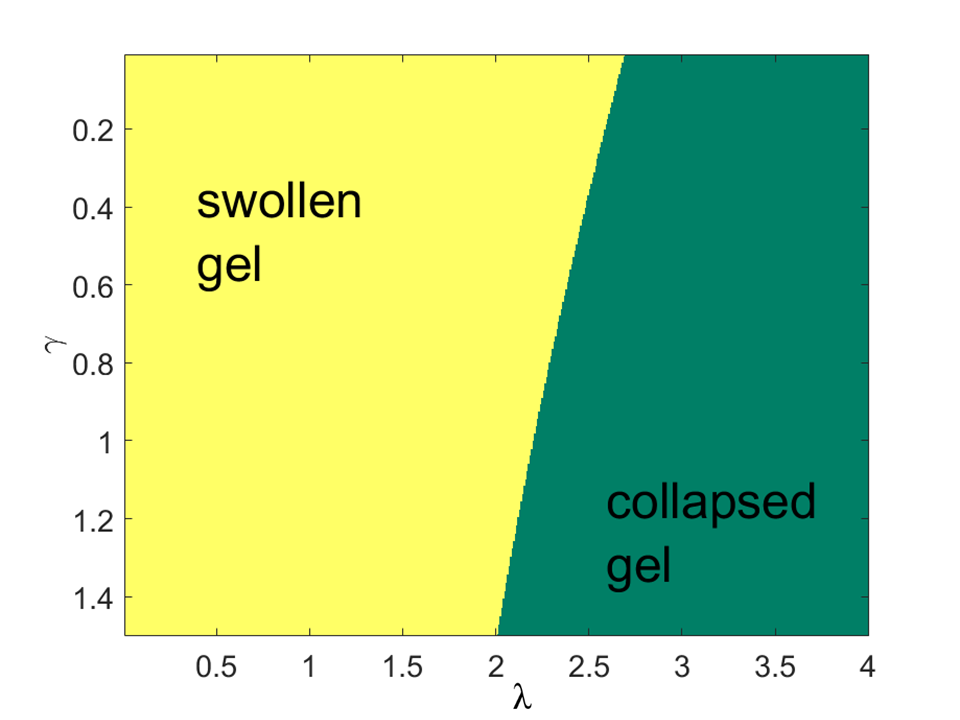}}
\caption{The phase diagram for the dipolar gel plotted in $\gamma$ {\sl vs} $\lambda$ coordinates. The data are shown for $\chi=0.25$, $v/l^3=1$, $\Phi_0=10^{-3}$, $m=500$.}
\label{Fig3}
\end{figure}

Further, we would like to discuss the dipole nonlocality effect on the jump-like collapse of the dipolar polymer gel. Fig. \ref{Fig3} demonstrates a phase diagram in the $\gamma$ {\sl vs} $\lambda$ coordinates. There is a certain curve separating two regions, where the gel is swollen (yellow) or collapsed (green). In accordance with Eq. (\ref{second_vir}), the account of dipole nonlocality (i.e. nonzero $\gamma$ values) shifts the coexistence between swollen and collapsed gel to the lower values of $\lambda$. We would like to note that the condition $\gamma \approx 1$ can be realized for zwitterionic polymers, such as polyesters, polyphosphazenes, and polyphosphobetaines \cite{kanduvc2020modeling}. For the polyelectrolyte chains, immersed in the low-polar solvents \cite{zaroslov1999change,khokhlov1994polyelectrolyte}, where the monomer units and counterions usually form solvent-separated ionic pairs with a fluctuating mutual distance, $\gamma$ can be greater than unity. It is also worth noting that at sufficiently small $\gamma$ ($l\ll b$), the contribution of the electrostatic interactions between the dipolar monomer units can be described with good accuracy by the electrostatic free energy of the freely moving dipolar particles (see, eq.(\ref{dipol})). Note that we have discussed this limiting regime earlier in the context of the conformational behavior of a single dipolar polymer chain with point-like dipoles on its monomer units \cite{budkov2017polymer}. However, for the subchains with $\gamma\sim 1$, the contribution (\ref{Fcor,ch}) to the electrostatic free energy becomes considerable. In other words, when the dipole length is comparable with the Kunh length, the approximation of disconnected monomers, used for different condensed polymer systems with the short-range volume interactions \cite{lifshitz1978some,khokhlov1994statistical}, gives a large discrepancy.    

Now we would like to discuss the parameters of the real polymers, for which one can expect to observe experimentally the discussed above transition. For instance, the dipole moment of monomer units of the polybetaines can reach $\approx 24-34~D$ (see \cite{kumar2009theory} and references therein). Thus, the dipole length, $l=p/e$, where $e$ is the elementary charge, for these polymers is in range $\approx 0.5-0.7~nm$. For the polymer chains with dipole moment $p=24~D$, immersed in water at $T=300~K$ ($\varepsilon\approx 78$) we obtain the coupling parameter $\lambda\approx 6$ which is in the region of the collapsed gel, depicted on Fig. \ref{Fig3} for $\Phi_0=10^{-3}$. In order to avoid the collapse, in this case, it is necessary to enhance significantly the reference volume fraction, $\Phi_0$, of the monomer units.

We would like to note that the obtained electrostatics driven gel collapse should not be confused with the co-nonsolvency transition taking place in polymer gels and solutions near low critical solution temperature  (LCST)\cite{budkov2018models,budkov2016statistical,budkov2017statistical,mukherji2019soft}. As is well known, in the zwitterionic hydrogels there are two types of phase transitions \cite{lei2018zwitterionic}. The first one is the gel collapse with the LCST, which takes place at a sufficiently small number of zwitterionic groups on the polymer chains. The gel collapse in this case proceeds with an increase in the temperature. The second transition with an upper critical solution temperature (UCST) is realized when the linear density of zwitterionic groups on the polymer backbone exceeds a certain threshold value. Such a dramatic transition, occurring upon cooling, may be caused by the dipole-dipole interactions of highly polar monomer units. Note also that we do not consider the poor solvent regime, where the volume attractive interactions lead to the gel collapse \cite{tanaka1978collapse,khokhlov1980swelling}. In this case, the additional dipole-dipole attractive interactions of monomer units just shift the transition temperature to higher values and the poor solvent concentration to lower values as in the mentioned above experiment \cite{zaroslov1999change}. In the present work we also do not consider the salt effect on the gel collapse. From general considerations, the addition of the salt ions into the gel will result in a screening of the dipole-dipole attractive interactions, thereby, preventing the collapse of the gel. The expansion of the dipolar macromolecules in solvent media with the increase of the ionic strength is known as the {\sl antipolyelectrolyte} effect \cite{kudaibergenov2006polymeric,kumar2009theory,xiang2018ionic}. An extensive theoretical study of the antipolyelectrolyte effect in the dipolar gels undoubtedly deserves to be published as separate research elsewhere. Finally, note that we neglect influence of the polymer on the dielectric constant of the gel (dielectric mismatch effect) \cite{budkov2017polymer,khokhlov1994polyelectrolyte,kumar2014enhanced}. This assumption can be justified by rather small volume fractions of the polymer, considered in present study. However, accounting for the dielectric mismatch will not qualitatively change system behavior, only slightly shifting the region of the gel collapse.

In conclusion, taking into account the conformational entropy of the polymer network within the independent subchains concept, excluded volume and dispersion interactions of the monomer units within the Flory-Huggins mean-field theory, and electrostatic interactions within the random phase approximation, we have developed a molecular theory of the liquid-phase dipolar polymer gels. We have established that for rather small volume fractions of the polymer, the presence of the dipole moments on the monomer units reduces the second virial coefficient. Namely, the dipole-dipole interactions can result in the negative second virial coefficient value even in the regime of good solvent. More importantly, the chain connectivity between the dipolar monomer units increases their effective attractive interactions. In addition, this phenomenon will be more pronounced in gels with the Kuhn lengths shorter than the dipole ones. We have found that sufficiently large electrostatic interactions between the monomer units can provoke the gel collapse in the regime of the good solvent. It proceeds like a first-order phase transition (an abrupt decrease in the expansion factor) at a sufficiently small polymer volume fraction. However, when latter exceeds a certain threshold value, the transition becomes smooth. We have demonstrated that the described phase transition can be realized at accessible physical parameters. Especially, we found that the predicted gel collapse resembles the one occurred in the zwitterionic hydrogels in the vicinity of the upper critical solution temperature \cite{lei2018zwitterionic}. We believe that the described phase transition could be used as an additional tool to control the swelling degree of gels in their different applications, such as molecular uptake, nanoreactors, coatings, membranes, artificial skin, toys production, {\sl etc}.

\section{Appendix}
Here we demonstrate a derivation of the screening function of the dipolar polymer gel within the random phase approximation (RPA) \cite{budkov2020statistical}. The screening function within such approach can be calculated as follows \cite{budkov2020statistical}
\begin{equation}
\label{screen}
\kappa^2({\bf k})=\frac{4\pi}{\varepsilon k_{B}T}C({\bf k}),
\end{equation}
where $C({\bf k})$ is the Fourier-image of the charge density correlation function
\begin{equation}
C({\bf r}-{\bf r}^{\prime})=\left<\hat{\rho}({\bf r})\hat{\rho}({\bf r}^{\prime})\right>,
\end{equation}
where $\left<(...)\right>$ means averaging over statistics of the polymer subchains without dipoles. Taking into account that the microscopic charge density is determined by the following expression
\begin{equation}
\hat{\rho}({\bf r})=q\sum\limits_{i=1}^{n}\sum\limits_{\alpha_{i}=1}^{m}\left(\delta({\bf r}-{\bf r}_{i}^{\alpha_{i}})-\delta({\bf r}-{\bf r}_{i}^{\alpha_{i}}-{\bf \xi}_{i}^{\alpha_{i}})\right),
\end{equation}
we obtain
\begin{equation}
C({\bf r}-{\bf r}^{\prime})=
q^2\sum\limits_{i,j}\sum\limits_{\alpha_{i},\gamma_{j}}\int\frac{d{\bf k}}{(2\pi)^3}\int\frac{d{\bf p}}{(2\pi)^3}\left<e^{-i{\bf k}{\bf r}_{i}^{\alpha_{i}}-i{\bf k}{\bf r}_{j}^{\gamma_{j}}}\right>_{r}\times
\end{equation}
\begin{equation}
\left<\left(e^{-i{\bf k}{\bf \xi}_{i}^{\alpha_{i}}}-1\right)\left(e^{-i{\bf p}{\bf\xi}_{j}^{\gamma_{j}}}-1\right)\right>_{\xi}e^{i{\bf k}{\bf r}+i{\bf p}{\bf r}^{\prime}},
\end{equation}
where indices $(i,j)$ enumerate the subchains, while $(\alpha_{i},\gamma_{j})$ -- their monomer units. We have also used the Fourier-representation of the Dirac delta-function
\begin{equation}
\delta({\bf x})=\int\frac{d{\bf k}}{(2\pi)^3} e^{i{\bf k}{\bf x}}.
\end{equation}

Using the identity \cite{budkov2020statistical}
\begin{equation}
\left<e^{-i{\bf k}{\bf r}_{i}^{\alpha_{i}}-i{\bf p}{\bf r}_{j}^{\gamma_{j}}}\right>_{r}=\frac{(2\pi)^3}{V}\delta({\bf k}+{\bf p})\left<e^{-i{\bf k}({\bf r}_{i}^{\alpha_{i}}-{\bf r}_{j}^{\gamma_{j}})}\right>_{r},
\end{equation}
we arrive at
\begin{equation}
\nonumber
C({\bf r}-{\bf r}^{\prime})=\frac{q^2}{V}\sum\limits_{i,j}\sum\limits_{\alpha_{i},\gamma_{j}}\int\frac{d{\bf k}}{(2\pi)^3}e^{i{\bf k}\left({\bf r}-{\bf r}^{\prime}\right)}\left<e^{-i{\bf k}({\bf r}_{i}^{\alpha_{i}}-{\bf r}_{j}^{\gamma_{j}})}\right>_{r}\times
\end{equation}
\begin{equation}
\left<\left(e^{-i{\bf k}{\bf \xi}_{i}^{\alpha_{i}}}-1\right)\left(e^{i{\bf k}{\bf \xi}_{j}^{\gamma_{j}}}-1\right)\right>_{\xi}=
\end{equation}
\begin{equation}
\nonumber
\frac{q^2n m}{V}\int\frac{d{\bf k}}{(2\pi)^3}e^{i{\bf k}\left({\bf r}-{\bf r}^{\prime}\right)}\left(2-\omega({\bf k})-\omega(-{\bf k})\right)+
\end{equation}
\begin{equation}
\frac{q^2}{V}\sum\limits_{i,j}\sum\limits_{\alpha_{i}\neq{\gamma}_i,\gamma_{j}}\int\frac{d{\bf k}}{(2\pi)^3}e^{i{\bf k}\left({\bf r}-{\bf r}^{\prime}\right)}\left<e^{-i{\bf k}({\bf r}_{i}^{\alpha_{i}}-{\bf r}_{j}^{\gamma_{j}})}\right>_{r}|1-\omega({\bf k})|^2,
\end{equation}
where we extracted averaging over the monomer units coordinates ${\bf r}_{i}^{\alpha_{i}}$ and displacements ${\bf \xi}_{i}^{\alpha_{i}}$ of the grafted charged centers and took into account that $\omega({\bf k})=\left<e^{i{\bf k}{\bf \xi}_{i}^{\alpha_{i}}}\right>_{\xi}$. Further, using the definition of the structure factor \cite{de1979scaling}
\begin{equation}
S({\bf k})=1+\frac{1}{n m}\sum\limits_{i,j}\sum\limits_{\alpha_{i}\neq \gamma_{i},\gamma_{j}}\left<e^{-i{\bf k}({\bf r}_{i}^{\alpha_{i}}-{\bf r}_{j}^{\gamma_{j}})}\right>_{r},
\end{equation}
we obtain
\begin{equation}
C({\bf r}-{\bf r}^{\prime})=\int\frac{d{\bf k}}{(2\pi)^3}e^{i{\bf k}\left({\bf r}-{\bf r}^{\prime}\right)}C({\bf k}),
\end{equation}
where
\begin{equation}
C({\bf k})=2q^2 c\left(1-Re(\omega({\bf k}))+\frac{1}{2}\left(S({\bf k})-1\right)|1-\omega({\bf k})|^2\right)
\end{equation}
with the monomer unit concentration $c=mn/V$. Therefore, according to eq. (\ref{screen}), the screening function takes the following form
\begin{equation}
\kappa^2({\bf k})=\frac{8\pi q^2c}{\varepsilon k_{B}T}\left(1-Re(\omega({\bf k}))+\frac{1}{2}\left(S({\bf k})-1\right)|1-\omega({\bf k})|^2\right),
\end{equation}
where $Re(\omega({\bf k}))$ is the real part of the complex function, $\omega({\bf k})$. In the case of a spherically symmetric distribution function, for which $\omega(-{\bf k})=\omega({\bf k})$ and $Re(\omega({\bf k}))=\omega({\bf k})$, we obtain eq. (5), written in the main text. In the absence of chain connectivity, when $S({\bf k})=1$, the screening function transforms into the expression for the solution of low-molecular weight dipolar molecules obtained earlier \cite{budkov2020statistical}. For the case of the unbound charged sites ($\omega({\bf k})=0$), the screening function transforms into the expression for salt-free polyelectrolyte solutions obtained for the first time in paper \cite{borue1988statistical}.

\section*{Acknowledgments}
This research was partially supported by grant 18-29-06008 from RFBR.

\scriptsize{
\bibliography{references} 
\bibliographystyle{rsc}} 

\end{document}